\documentclass[fleqn,10pt]{wlscirep}
\title{Inside the perpendicular spin-torque memristor}

\author[1,*]{Steven Lequeux}
\author[1,2]{Joao Sampaio}
\author[1]{Vincent Cros}
\author[3]{Kay Yakushiji}
\author[3]{Akio Fukushima}
\author[3]{Rie Matsumoto}
\author[3]{Hitoshi Kubota}
\author[3]{Shinji Yuasa}
\author[1]{Julie Grollier}
\affil[1]{Unit\'e Mixte de Physique, CNRS, Thales, Univ. Paris-Sud, Université Paris-Saclay, 91767, Palaiseau, France.}
\affil[2]{Laboratoire de Physique des Solides, CNRS, Univ. Paris-Sud, Université Paris-Saclay, 91405, Orsay , France.}
\affil[3]{National Institute of Advanced Industrial Science and Technology (AIST), Tsukuba, Ibaraki 305-8568, Japan.}

\affil[*]{steven.lequeux@thalesgroup.com}

\begin{abstract}
Memristors are non-volatile nano-resistors. Their resistance can be tuned by applied currents or voltages and set to a large number of levels between two limit values. Thanks to these properties, memristors are ideal building blocks for a number of applications such as multilevel non-volatile memories and artificial nano-synapses, which are the focus of this work. A key point towards the development of large scale memristive neuromorphic hardware is to build these neural networks with a memristor technology compatible with the best candidates for the future mainstream non-volatile memories. Here we show the first experimental achievement of a memristor compatible with Spin-Torque Magnetic Random Access Memory. The resistive switching in our spin-torque memristor is linked to the displacement of a magnetic domain wall by spin-torques in a perpendicularly magnetized magnetic tunnel junction. We demonstrate that our magnetic synapse has a large number of intermediate resistance states, sufficient for neural computation. Moreover, we show that engineering the device geometry allows leveraging the most efficient spin torque to displace the magnetic domain wall at low current densities and thus to minimize the energy cost of our memristor. Our results pave the way for spin-torque based analog magnetic neural computation.

\end{abstract}
\begin{document}

\flushbottom
\maketitle
%
%
\thispagestyle{empty}

\section*{Introduction}

The resistance of a memristor depends on the amplitude and duration of the successive voltages that have been previously applied between its electrodes  \cite{Chua1971}. The resulting intricate resistance versus voltage hysteresis loops illustrated in Fig. \ref{fig1}a are the hallmark of memristor nanodevices. Thanks to their memory and tunability features, memristors can imitate a very important property of biological synapses: their plasticity \cite{Jo2010}. Synaptic plasticity is the ability of synapses to reconfigure the strength with which they connect two neurons according to the past electrical activity of these neurons. In the brain, this phenomenon allows memories to be formed and stored. In hardware, artificial synapses endowed with plasticity can allow neural networks to learn and adapt to a changing environment. Today, memristors are considered as the best solution to realize such plastic nanosynapses. Fabricating memristor devices requires triggering voltage induced resistive changes at the nanoscale. Several physical effects can be harnessed for this purpose: reduction-oxydation phenomena \cite{Strukov2008}, phase changes \cite{Kuzum2012, Suri2012}, atomic displacements \cite{Hasegawa2010} or ferroelectric switching \cite{Chanthbouala2012}. The current challenge is to go beyond the realization of single devices and to build dense and functional arrays of memristors. Given the spread of possible solutions, all of them with their respective advantages and downsides \cite{Yang2013}, it is likely that future neuromorphic hardware will be built using the non-volatile memory technology that will be available on the market in the next few years. While competition is fierce in this domain, thanks to recent advances, the Spin-Torque Magnetic Random Access Memory is among the most promising technologies \cite{Lee2014}.

In this report, we show the first experimental realization of a spin-torque memristor with the potential to provide massive local memory access to large scale magnetic neural networks. As illustrated in Fig. \ref{fig1}b, the spin-torque memristor is a magnetic tunnel junction with a single magnetic domain wall in its free layer. Thanks to the spin-torque effect, the domain wall (DW) can be displaced back and forth by applying positive or negative voltages across the stripe-shaped junction. When the voltage is removed, the domain wall stabilizes in different pinning sites corresponding to different positions along the track. Each domain wall position has a different resistance values thanks to the tunnel magnetoresistance effect, leading to the desired memristive features. 

Based on a magnetic tunnel junction, which is also the building block of the next generation of magnetic memories, our spin-torque memristor has the advantages of this technology: CMOS compatibility, high cyclability and fast switching \cite{Lee2014, Ikeda2007}. As shown in Fig. \ref{fig1}b, in our spin-torque memristor device, the currents for reading and writing the resistance states are both injected perpendicularly to the magnetic layers. Indeed two-terminal devices are most promising for miniaturization and integration in neuromorphic networks \cite{Indiveri2013}. In addition, in this geometry, the currents incoming from afferent neurons and flowing through the junction are not only modulated by the junction resistance, but can also modify the junction resistance depending on their respective amplitudes and durations. In this way, the transmission of two terminal artificial synapses can evolve autonomously according to local voltages, allowing unsupervised learning \cite{Kim2012,Basheer2000}. These considerations justify the use of classical spin torque over spin-orbit torques, which require a three-terminal geometry \cite{Mihai2011a, Cubukcu2014, Fukami2016}. Moreover, we have shown recently that the classical spin torque can displace magnetic domain walls with low current densities (about 10$^{6}$ A/cm$^{2}$ \cite{Chanthbouala2011}) and large speeds (over 500 m/s \cite{Sampaio2013}) when the spin-polarized current is injected vertically in magnetic tunnel junctions. Here, we show that we can realize a spin-torque memristor with a large number of resistance states by using a magnetic tunnel junction with perpendicular magnetic anisotropy. Moreover, we show that by carefully engineering our system we can use the largest spin-torque to drive the domain wall, leading to low current density operation. 

\section*{Results}

\begin{figure}[ht]
   \centering
	\includegraphics[width=\textwidth]{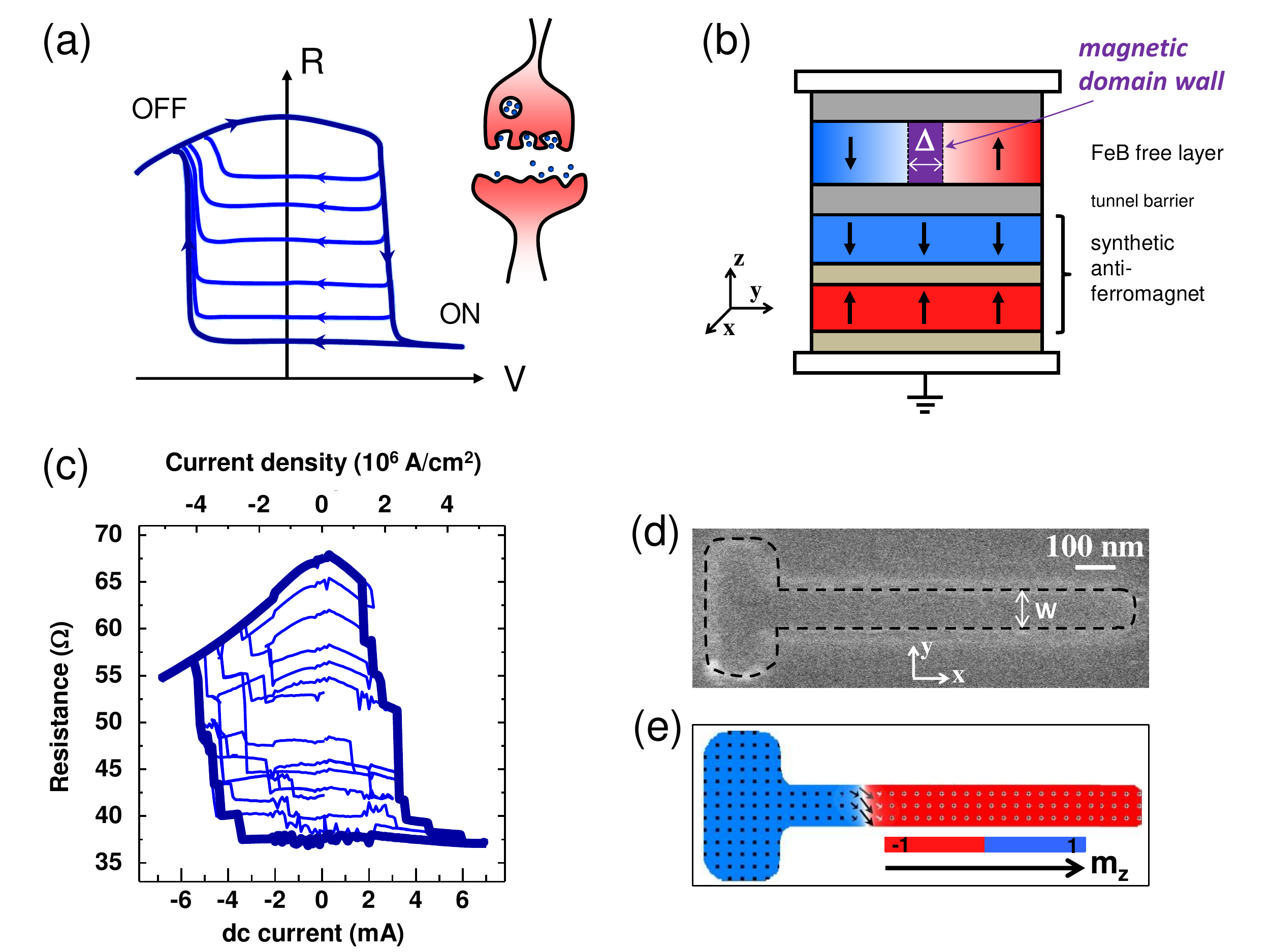}
     \caption{Spin-torque memristor. (a) Typical resistance versus voltage cycles characteristic of a memristor. The red dots represent the intermediate resistance state at zero bias. Inset: sketch of a biological synapse. (b) Schematic of our MgO based magnetic tunnel junction with the domain wall in the FeB free layer. $\Delta$ is the width of the domain wall. (c) Resistance as a function of the vertically injected dc current (sweeped in the same direction than the one shown by the arrows in Fig. \ref{fig1} (a)), measured at an external field $H_z$=85 Oe. (d) Scanning electron microscope image of the sample, with a black dashed line to emphasize its contour. (e) Micromagnetic simulations of the domain wall propagating in a magnetic track of 100 nm width. The tilt angle of the spin in the domain wall structure shows that for this width, the domain wall is hybrid between Neel and Bloch configurations.}
\label{fig1}
\end{figure}

The magnetic tunnel junctions are fabricated by electron beam lithography and ion beam milling from a film deposited by magnetron sputtering. A side view of the  samples is given in Fig. \ref{fig1}b, where the domain wall is the purple region of width $\Delta$. The magnetic stack consists of a synthetic anti-ferromagnet [CoPt(2.4) / Ru(0.9) / CoPt(1.3) / Ta(0.2) / FeB(1)], a tunnel barrier MgO(1), a magnetic free layer [FeB(1.2) / Ta(0.3) / FeB(0.7)], and a capping layer [MgO(1) / Ta(6)], with thicknesses in nanometers \cite{Yakushiji2013}. The tunnel magneto-resistance ratio is about 95 \% at room temperature for a minimum resistance of 36 $\Omega$. In the case of magnetic tunnel junctions, the two extreme values of resistance in Fig. \ref{fig1}a correspond to the parallel (P) and antiparallel (AP) magnetizations configurations of the free layer and the top layer of the synthetic anti-ferromagnet. A top view of the samples is shown in Fig. \ref{fig1}d. The domain wall propagates in a micrometer-long track of width $W$. The samples also include a nucleation pad with larger dimensions than the track designed to facilitate the creation of the domain wall in the free layer. The samples are fabricated with three different track widths $W$ of 90, 100 and 110 nm, and with three different sizes of nucleation pad.

Fig. \ref{fig1}c shows the evolution of the junction resistance measured as a function of the vertically injected dc current. Positive currents correspond to an electron flow from the synthetic anti-ferromagnet to the free layer. The measurement is performed in the presence of an external field $H_z$=85 Oe applied along the perpendicular direction in order to suppress the stray field emitted by the synthetic anti-ferromagnet. Starting from a uniform state (P or AP), a domain wall is nucleated by vertical injection of current. Then, it can be displaced back and forth, towards the AP state with negative currents, and towards the P state with positive currents. During its propagation, the domain wall gets pinned in random defects at the edges or the surface of the sample. This gives rise to the multiple levels of resistance observable in Fig. \ref{fig1}c. Note that the random quantization of resistance in our memristor is not an issue for synaptic operation \cite{Querlioz2013}. For each of the five samples we have measured under dc current injection, the number of observed intermediate resistance states is comprised between 15 and 20.  D. Querlioz \textit{et al.} have simulated a neuromorphic system where each synapse is composed of several binary magnetic tunnel junctions in parallel \cite{Querlioz2013, Querlioz2015}. They show (see Fig. 6 of Ref\cite{Querlioz2015}) that 7 to 10 binary junctions are necessary in each synapse for recognizing handwritten digits with a reasonable rate above 75 \%. This result, transposed to our system, means that 13 to 19 resistance levels should be sufficient for pattern classification tasks. In addition, the number of resistance levels can be increased in the future by applying voltage pulses of a few nanoseconds leading to smaller domain wall displacements than dc currents. Indeed, at the end of the pulse, the domain wall will be able to stop in defects that would otherwise be too weak to pin it when pushed by a constant current. These results indicate that the spin-torque memristor is appropriate for neuromorphic applications.

\begin{figure}[ht]
   \centering
	\includegraphics[width=0.8\textwidth]{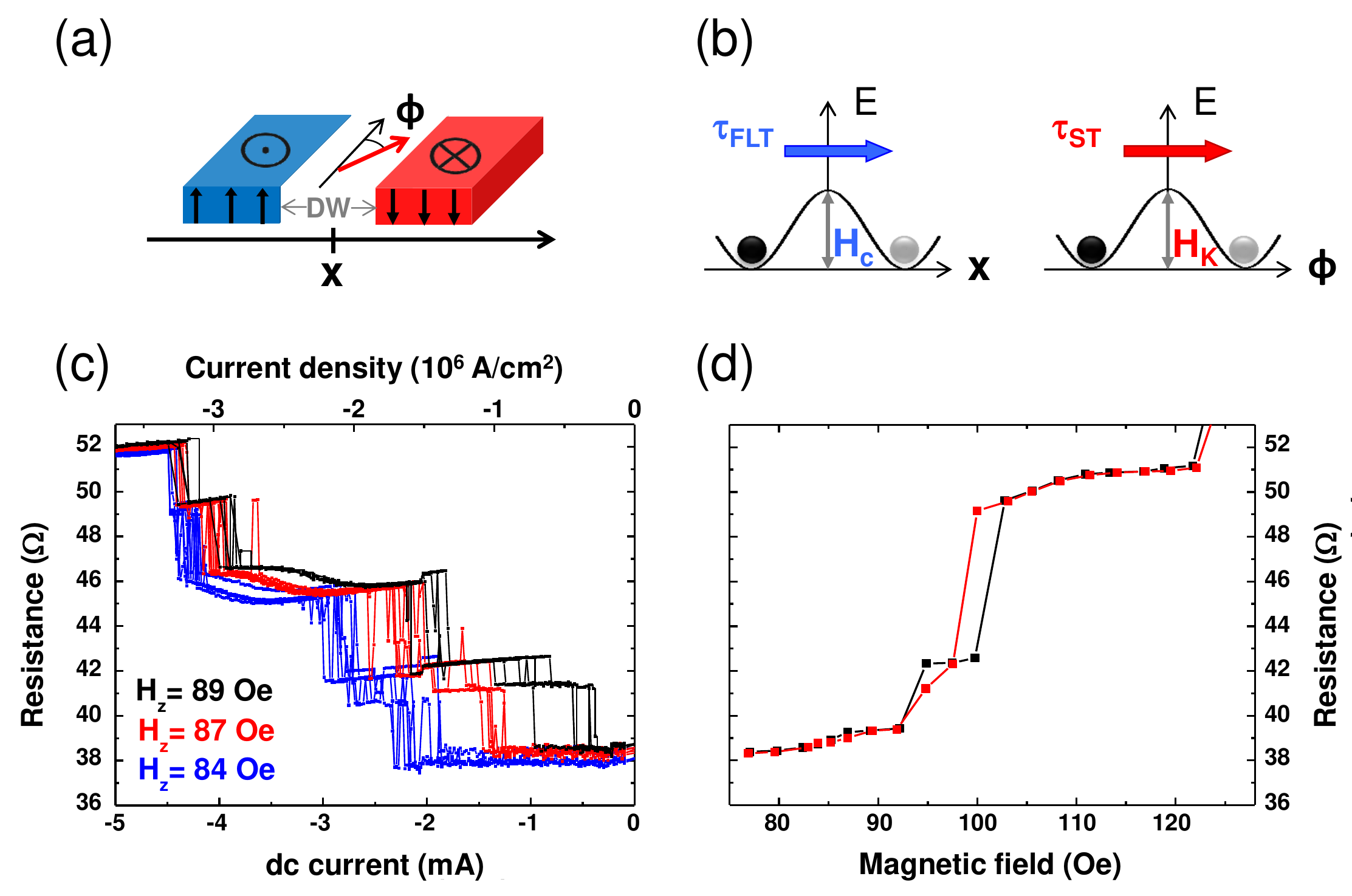}
     \caption{Action of the different spin-torques on the magnetic domain wall. (a) Illustration of the domain wall and its coordinates in a track. The domain wall position is defined by $x$, $\phi$ is its internal tilt angle. (b) Schematic representation of the energy landscape for a domain wall depinned by the Field-like-torque (left) and the Slonczewski torque (right). $H_c$ is the depinning field intrinsic to the pinning center, and $H_K$ is the domain wall anisotropy field. (c) Resistive changes during current driven domain wall depinning from a given pinning center (resistance value of 38 $\Omega$) for different values of the external magnetic field in the range [78-92] Oe. (d) Resistive changes during field (parallel to the z-axis) driven domain wall depinning from the same pinning center. The measurement has be done twice. }
\label{fig2}
\end{figure}

Additionally to the large number of intermediate states, the resistance versus current loops of our memristors, displayed in Fig. \ref{fig1}c, always show two notable features. First, domain wall displacements are obtained for current densities of the order of 10$^6$ A/cm$^2$. This value is remarkably small for spin-torque induced domain wall motion in perpendicularly magnetized materials \cite{Boulle2008a,Ravelosona2007, Moore2008, Fukami2009}. Secondly, the switching between different resistance levels appears highly stochastic. Figs. \ref{fig2}c and \ref{fig2}d show the typical experimental resistive variations associated with domain wall depinning in our samples of width 100 nm. At zero current the domain wall is in a pinning site corresponding to a resistance value of 38.5 $\Omega$. Fig. \ref{fig2}c shows domain wall depinning by current for several values of the external field. In that case, the domain wall motion is extremely stochastic close to the threshold current values for depinning, jumping back and forth between pinning centers separated by several hundreds of nanometers. These stochastic features, that arise when the domain wall is pushed by current, are a potential asset for neural computation \cite{Aldo2008,Querlioz2015}. When the domain wall depinning is driven by a magnetic field instead of a current, the stochasticity completely disappears, as can be seen in Fig. \ref{fig2}d. This means that spin-torques and magnetic fields exert completely different forces on the domain wall. In the following, we describe how we have engineered our magnetic tunnel junctions in order to obtain these high efficiency and stochasticity of spin-torque driven domain wall motion.

When the current is injected vertically in a magnetic tunnel junction, two spin torques are exerted on the magnetization of the free layer: the Slonczewski torque (ST) and the Field-like-torque (FLT) \cite{Slonczewski1989}. The Field-like-torque is smaller than the Slonczewski torque, and the ratio of their amplitudes, $\zeta$, is typically 30\% in magnetic tunnel junctions \cite{Kubota2007}. We have designed our magnetic tunnel junctions in order to leverage the largest spin-torque, the Slonczewski torque, to displace the domain wall. Indeed, the Slonczewski torque and the Field-like-torque have very different actions on the domain wall depinning process. Fig. \ref{fig2}b illustrates the energy landscape in each case. On the left is sketched the case of domain wall depinning governed by the Field-like-torque only. In our geometry, the Field-like-torque is equivalent to a field applied in the z direction, along the spins in the domains on each side of the wall. Therefore it has the proper symmetry to push the domain wall along the track (along the x axis in Fig. \ref{fig2}d), by modifying its position $x$. The energy barrier that the Field-like-torque has to overcome in order to move the domain wall is directly related to the depinning field $H_c$, an intrinsic property of the pinning center. The right part of Fig. \ref{fig2}b shows the case of a domain wall depinning governed by the Slonczewski torque only. By symmetry this torque can modify the tilt angle $\phi$ of the magnetization within the domain wall (Fig. \ref{fig2}d). Therefore the Slonczewski torque can depin a domain wall by rotating its internal angle. The energy it has to overcome is the domain wall anisotropy $K$. Once the domain wall is depinned the Slonczewski torque continues to increase its $\phi$ angle. This drives the domain into a precessional regime known as the Walker breakdown, where the domain wall continuously oscillates between the Bloch and N\'eel configurations \cite{Sampaio2013,Thiaville2006a}. Making the Slonczewski torque the most efficient spin-torque to depin the domain wall requires lowering the domain wall anisotropy field $H_K$ below the coercive field $H_c$. One way to reach this condition is to design an anisotropy-less domain wall, that can freely oscillate between the Bloch and N\'eel configurations without having to overcome an energy barrier. Such an hybrid domain wall can be obtained by choosing the stripe width carefully. For large stripe widths, the spins inside the domain wall can point in the direction perpendicular to the stripe, and the Bloch configuration is favoured. For small stripe widths, the transverse demagnetizing field increases and the spins inside the domain wall tend to point along the stripe, resulting in a N\'eel configuration. For intermediate widths at the frontier between these two regimes, the domain wall is hybrid and can easily rotate between the N\'eel and Bloch magnetic configurations \cite{Koyama2011, DeJong2015}. 

In order to find the optimal track width $W$ leading to a hybrid domain wall and low current-density propagation through the Slonczewski torque, we have performed micromagnetic simulations, using material parameters determined experimentally \cite{Yakushiji2013, Konoto2013} (see methods). We found that values of $W$ in the range [90-110] nm allow creating an hybrid domain wall in the stripe. Fig. \ref{fig1}e shows the result of micromagnetic simulations for a 100 nm width stripe. As can be seen, the angle of spins inside of the domain wall is close to 45 $\deg$, in between the N\'eel and Bloch configurations as expected from an hybrid domain wall. The large difference between spin-torque and field driven domain wall motion observed experimentally in Figs. \ref{fig2}c and \ref{fig2}d shows that our optimization was successful and that the spin-torque responsible for domain wall motion is not a Field-like torque, but rather the Slonczewski torque. In addition, the enhanced mobility of the domain wall under current, as well as its stochastic back and forth motion, indicates that we have, as desired, created a hybrid domain wall propagating in the Walker regime when it is set in precession by the Slonczewski torque \cite{Jiang2010}.

\section*{Discussion}

In order to quantify the contribution of each spin torque to the domain wall motion in our samples, we numerically compute the evolution of the depinning current density  $J_{th}$ as a function of magnetic field, and compare it to our experiments. Indeed, depending on which torque drives the depinning, the variation of $J_{th}$ with field should be radically different. If the Field-like-torque dominates the depinning process, the effect of current is equivalent to an applied magnetic field in the z direction, and $J_{th}$ should vary largely with the external field \cite{Chanthbouala2011}. On the contrary, if the Slonczewski torque drives the domain wall, the energy barrier, set by the domain wall anisotropy $K$, is independent of the applied field, and $J_{th}$ should depend weakly on the applied field. We simulate numerically the domain wall depinning process under each torque by solving the following system of equations \cite{Thiaville2006a} describing a 1D translational domain wall motion:

\begin{align}
  \begin{cases}
\dot{\phi} + \alpha \frac{\dot{x}}{\Delta} = \gamma H_{z} + \gamma \sigma_{FLT} - \gamma \frac{x}{x_{c}} H_{c}  \\
\frac{\dot{x}}{\Delta} - \alpha \dot{\phi}   = \gamma \sigma_{ST} + \frac{1}{2} \gamma H_{K} \sin(2\phi)
  \end{cases}
\label{eq1}
\end{align}

\begin{figure}[ht]
   \centering
	\includegraphics[width=0.7\textwidth]{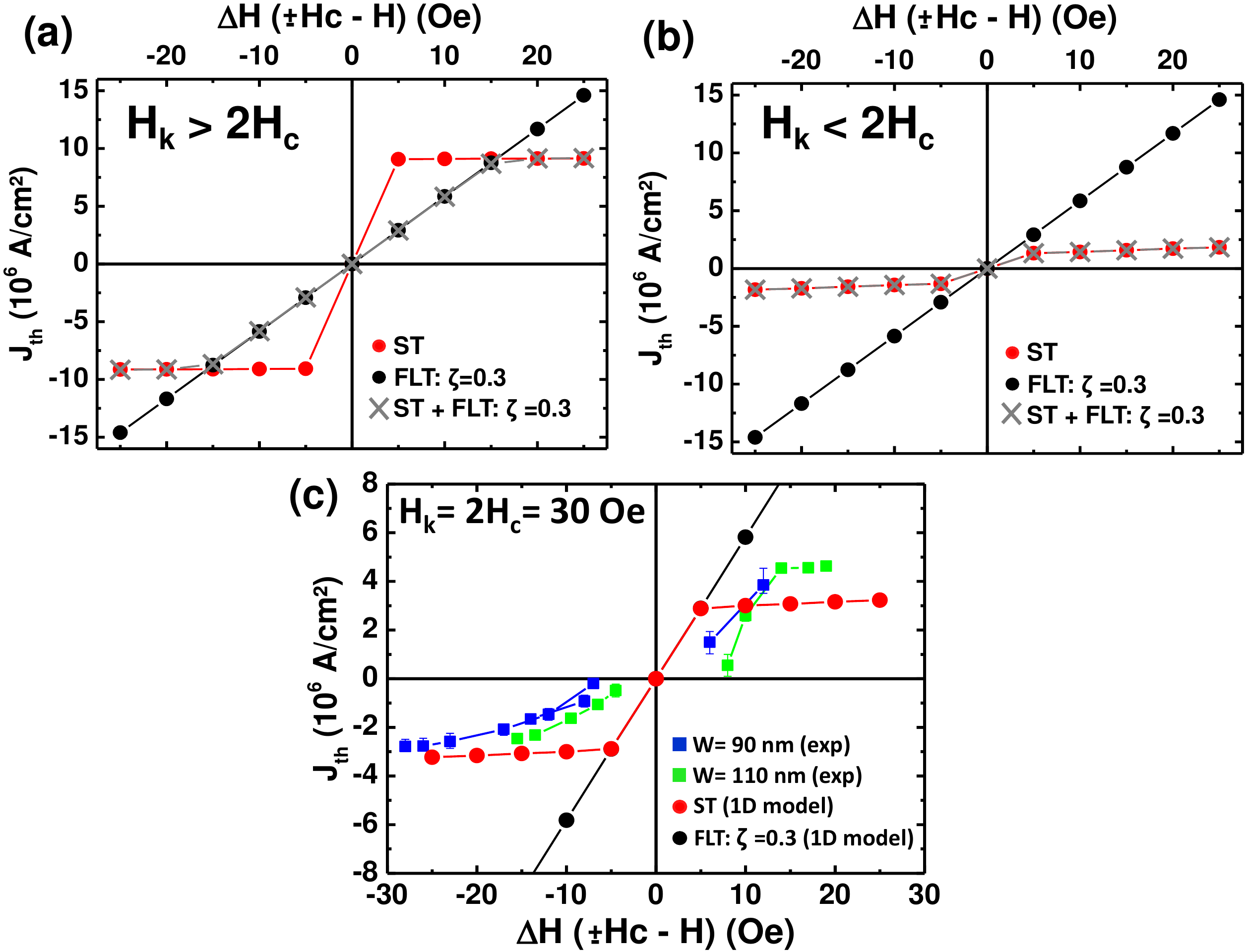}
     \caption{Depinning threshold current as a function of magnetic field. (a),(b) Numerical calculation of the depinning threshold current density $J_{th}$  as a function of $\Delta H$ for $H_{K}>$$H_{c}$ (a) and $H_{K}<$$H_{c}$ (b). Different combinations of spin torques are considered. The red and black circles show respectively the evolution of $J_{th}$ when the depinning process is governed only by the Slonczewski torque and only by the Field-like torque, with a ratio FLT/ST $\zeta$=0.3. The gray crosses show the case where both torques are considered together, still with $\zeta$=0.3. (c) Experimental measurements of $J_{th}$ for two different stripe widths $W$=90 nm (blue squares) and $W$=110 nm (green squares) compared with numerical calculations for $H_{K}=2$$H_{c}$=30 Oe.}
\label{fig3}
\end{figure}

In these equations, $x$ is the domain wall position and $\phi$ its internal angle; $\gamma$, $\alpha$ and $x_{c}$ represent the gyromagnetic ratio, the intrinsic damping constant, and the spatial extension of the potential well along the x axis; $H_{c}$ is the coercive field and $H_{K}$ the domain wall anisotropy field. $\sigma_{ST}$ and $\sigma_{FLT}$ represent the amplitudes of the Slonczewski and Field-like torques in units of magnetic field. This model accounts for the domain wall dynamics through the two variables $x$ and $\phi$, under the effects of spin torques ($\sigma_{FLT}$ and $\sigma_{ST}$), external field ($H_{z}$) and pinning potential ($x_{c}$ and $H_{c}$). In our simulations, we vary the value of the domain wall anisotropy field $H_{K}$, and set all the other parameters to values we have determined experimentally or through micromagnetic simulations (see methods). In particular, the coercive field  $H_c$ is taken to 15 Oe as in experiments. All details of numerical calculations are given in the Methods section. Fig. \ref{fig3} shows the computed variation of $J_{th}$ with $\Delta H$, defined as the difference between $H_{z}$ and $H_{c}$. Two cases are analyzed, $H_{K}>$$H_{c}$ ($H_{K}$=100 Oe in Fig. \ref{fig3}a), and  $H_{K}<$$H_{c}$ ($H_{K}$=10 Oe in Fig. \ref{fig3}b). For these two cases, we show the results obtained with the Slonczewski torque alone, the Field-like torque alone, and with both torques.

In agreement with Koyama et al. \cite{Koyama2011}, $J_{th}$ presents a weak dependence on $\Delta H$ when only the Slonczewski torque is considered (red circles in Figs. \ref{fig3}a and \ref{fig3}b). In the same way, a linear dependence is observed when only the Field-like torque is considered (black circles in Figs. \ref{fig3}a and \ref{fig3}b). When the two torques are acting together, in the case $H_{K}>$2$H_{c}$ we can notice that $J_{th}$ evolves in two different regimes. For the lowest values of $\Delta H$, below 15 Oe, it follows the same trend as in the case where the Field-like torque acts alone (gray crosses in Fig. \ref{fig3}a). For larger $\Delta H$, above 20 Oe, it remains constant, as in the case where only the Slonczewski torque is considered. This means that when the two torques are acting together, for a fixed value of $\Delta H$ (i.e. of the external field $H_z$), only one torque governs the domain wall depinning process. On the contrary, if $H_{K}<$2$H_{c}$, using the two torques together is identical to using the Slonczewski torque alone (respectively gray crosses and red circles in Fig. \ref{fig3}b). The Field-like torque has no additional effect on $J_{th}$. The relative values of $H_K$ and 2$H_c$ determine the dominant torque in the domain wall depinning process. In addition, in the case of a depinning governed only by the Slonczewski torque, the value of $H_K$ also determines the order of magnitude of $J_{th}$. The larger $H_K$, the higher the energy barrier to overcome, the larger the current densities $J_{th}$ necessary for depinning the domain wall.

Fig. \ref{fig3}c shows the experimental measurements of domain wall depinning. $J_{th}$ is plotted as a function of $\Delta H$ for two pinning sites obtained in different stripes of width $W$=90 nm (blue squares in Fig. \ref{fig3}c) and $W$=110 nm (green squares in Fig. \ref{fig3}c). The details of the measurements are given in the methods section. Experimentally, the maximum values of current densities needed to depin the domain wall are lower than 5.10$^{6}$ A/cm$^{2}$ for both stripe widths. We find that the best value of anisotropy field $H_K$ describing our experimental results is $H_{K}=2$$H_{c}$=30 Oe. The results of simulations with this value of $H_K$ are shown by circles in Fig. \ref{fig3}c. Clearly, the experimental curves are in better agreement with the red circles, which correspond to a domain wall depinning process governed by the Slonczewski torque rather than with the black circles corresponding to a domain wall depinning by the Field-like-torque. The slight asymmetry of the experimental curve can be due to non-parabolic potential wells or non-linear bias-dependences of the Slonczewski torque \cite{Chanthbouala2011}. The qualitative and quantitative correspondence between numerical simulations and experiments indicates that in our experiments the Slonczewski torque is dominant in the domain wall depinning process.

To summarize, we have shown for the first time the realization of a spin-torque memristor compatible with the future spin-torque magnetic memories. Our memristor has a large number of resistance states under dc current, between 15 and 20, which is sufficient to use it as an artificial nanosynapse. The current densities needed to drive the domain wall are low, of the order of 10$^{6}$ A/cm$^{2}$. Comparing the field dependence of depinning current densities with numerical simulations indicates that the largest spin-torque, the Slonczewski torque, drives the  domain wall motion in our samples, also implying that we have successfully engineered our devices to accommodate hybrid domain walls. Our results are a first important step towards the realization of analog bio-inspired magnetic computing hardware leveraging spin-torque induced switching in multi-level magnetic tunnel junctions for learning.

\section*{Methods}

\subsection*{Experimental measurements}
To probe the intermediate resistance states available for the DW (Fig. \ref{fig1}c), we have applied a fixed magnetic field of 85 Oe along the z direction. Starting first from the AP state, the DW is nucleated by the vertical injection of the dc current. Then, we have repeated several current cycles in the range [-6.5, 6.5] mA to displace the DW and probe its different available pinning centers. To study the evolution of $J_{th}$ as a function of the external field, we choose to focus on the two extreme values of stripe width W (90 and 110 nm, respectively blue and green lines in Fig. \ref{fig3}c). We have found a reproducible and stable pinning center in the two samples. For $W$=90 nm, the pinning center is stable on the field range: [92,118] Oe, and corresponds to a resistance level of 41 $\Omega$. In order to study the threshold depinning current $J_{th}$, we swept the dc current from 0 until the DW depinning by step of 20 $\mu$A. This process was repeated between 8 and 10 times for each values of the external field in the stability range of the probed pinning center. Exactly the same procedure was used for the stripe width of 110 nm, where the pinning center, associated to a resistance level of 38.5 $\Omega$, is stable in the field range [75,92] Oe (Fig. \ref{fig2}d). The stochastic behavior shown in Fig. \ref{fig2}c comes from the $J_{th}$ measurements for the 110 nm stripe width.\\

\subsection*{Numerical calculations}
Micromagnetic simulations were performed with the OOMMF code \cite{OOMMF}. For the numerical integration of the 1D model (Eq. \ref{eq1}), we used a 4th order Runge-Kutta method (RK4) with a step time of 5.10$^{-12}$ s (5 ps). In order to study the threshold current densities ($J_{th}$), we increased the simulated current density value using a ramp. The ramp rate was chosen to 10$^{4}$ A/cm$^{2}$/ ns, and was verified to have no impact on the results at zero temperature. The amplitude of the spin transfer torques $\sigma_{ST}$ and $\sigma_{FLT}$ were described as follows: $\sigma_{ST}=\frac{J P g \mu_{B}}{2 e Ms}$ and  $ \sigma_{FLT}= \zeta \sigma_{ST}$, where $\zeta$, the amplitude of the Field-like torque, is always of 0.3 in Fig. \ref{fig3}. In these expressions $J$ is the current density, and $g$, $\mu_{B}$ and $e$ are respectively the Lande factor, the magnetic permeability and the elementary charge. In order to estimate the spin polarization $P$ in our samples, we first calculated the spin polarization $P_{J}$ from Julliere's model: $TMR=\frac{2 P_{J}^{2}}{(1 - P_{J}^{2})}$. The two measured samples in Fig. \ref{fig3}c present a TMR ratio of 82 \% and 93 \% (respectively for $W$=90 and 110 nm), which correspond to $P_{J}= 0.55$ $\pm$ 0.01. J.C. Slonczewski and J.Z. Sun \cite{Slonczewski2007, Sun2008} have shown that the polarization factor $P$ in the expression of the spin torques efficiencies $\sigma_{ST}$ and $\sigma_{FLT}$ can be calculated as: $P=\frac{P_{J}}{(1 + P_{J}^{2})}$. Using this expression, we found a spin polarization factor $P$ of 0.422 $\pm$ 0.004.  The other material parameters were determined as follows. The damping constant $\alpha$ of 0.005 was measured for our samples stack in ref. \cite{Konoto2013}. The saturation magnetization value ($M_{s}$) of 1.05 10$^{6}$ A/m was obtained from magnetization versus field measurements \cite{Yakushiji2013}. The value of the perpendicular anisotropy of 7.10$^{6}$ erg/m$^{3}$, used to obtain the micromagnetic simulation result shown in Fig. \ref{fig1}e, was estimated from the same magnetization versus field measurements, where the field is applied in the plane of the free layer. The stripe's thickness is 2.2 nm. The pinning strength is described by two values: the spatial extension of the potential well along the x axis ($x_{c}$=100 nm), and the depinning (or coercive) field $H_{c}$ (15 Oe as in experiments). 
The width of the DW, 16.6 nm, was extracted from the magnetization profile of the z component along the x axis of the micromagnetic simulation result shown on Fig. \ref{fig1}e.


\section*{Acknowledgements}

The authors acknowledge financial support from the European Research Council (Starting Independent Researcher Grant No.~ERC 2010 Stg 259068) and the French ministry of defense (DGA).

\section*{Author contributions statement}

J.G. conceived the experiments; S.L, J.S. and R.M. performed numerical simulations; K.Y, A.F, H.K and S.Y. designed the junction stack and fabricated the samples;  S.L and J.S. conducted the experiments; S.L, J.S., V.C and J.G. analyzed the results.  All authors wrote and reviewed the manuscript.

\section*{Additional information}

The authors declare no competing financial interests.

\newpage

\section*{Figure captions}

\end{document}